\documentclass[12pt]{article}
\usepackage{graphicx}
\usepackage{epsfig}

\newcommand {\be} {\begin{equation}}
\newcommand {\ee} {\end{equation}}
\newcommand {\bea} {\begin{eqnarray} }
\newcommand {\eea} {\nonumber \end{eqnarray}}

\newcommand {\bc} {\begin{center}}
\newcommand {\ec} {\end{center}}
\newcommand {\bd}{\begin{displaymath}}
\newcommand {\ed}{\end{displaymath}}

\def \form#1 {eq. (\ref{#1}) }
\def \parziale#1#2  {{\partial {#1} \over \partial {#2}}}
\begin{document}
\title{Planck's Legacy to Statistical Mechanics}
\author{Giorgio Parisi \\ Dipartimento di Fisica, Sezione INFN and unit\`a INFM,\\
Universit\`a di Roma ``La Sapienza'',\\
Piazzale Aldo Moro 2,
I-00185 Roma (Italy)\\
Giorgio.Parisi@Roma1.INFN.it} 
\maketitle \abstract{In this talk I will describe the deep influence 
Planck had on the development of statistical mechanics.  At this end I will first outline the 
theoretical situation of statistical mechanics before Planck.  I will then describe his main 
contributions to this field and the progresses obtained as an immediate consequence of his work.  I 
will also outline the later evolution of statistical mechanics in relation with Planck's work.  I 
will finally report on a still unsolved problem in statistical mechanics,  historically 
related to the properties of black-body radiation.  }

\section{Introduction}
In this talk I will try to highlight the role that statistical mechanics played on Planck's discovery 
and the  consequences of Planck's discovery on statistical mechanics.
For reasons of time I cannot present a complete historical account \cite{KP} and I will select the 
points that seem to me more important.

I will first outline the situation in statistical physics at the end of the 19$^{th}$ century.  I 
will then present Planck's approach to radiation and his two first derivations (1900) of the famous 
formula for the black body radiation.  I will then outline the short term consequences of Planck's 
discovery on statistical mechanics and I will recall a part of the discussions  done after it was 
realized that it was impossible to derive Planck formula in equilibrium statistical mechanics.  I 
will go on by describing the discovery of quantum statistics and I will finally report on our 
present understanding of approach to equilibrium in weakly non linear systems, a problem that was 
already discussed at Planck's time and was relevant in the debate on the black body radiation.  
Finally in the conclusions I will present some personal considerations.

\section{Statistical Physics at the end of the 19$^{th}$ century}

At the end of the 19$^{th}$ century thermodynamics was well established. One of the most impressive 
results is the purely thermodynamical derivation of Wien's displacement law for the black body 
radiation using as input the first two laws of thermodynamics. 
If we call $E(\nu,T)$ the intensity of the black body radiation at frequency $\nu$, Wien's 
displacement law \cite{WDISP} states that

\be
E(\nu,T)=\nu^{3} \phi(\nu/T) \ , \label{WIEN}
\ee
where $\phi(\nu/T)$ is a function that cannot be determined on purely thermodynamics grounds.  The 
proof, which I have not checked, is rather sophisticated; it involves monochromatic filters, the 
analysis of the effect of the radiation pressure on a moving piston and the variation of the 
radiation spectrum as a consequence of the movement of the piston, using the Doppler effect.

However we could say that at that time there was no statistical mechanics.  The name was practically  
not present in the literature: what really existed was ``kinetic gas theory.''  Statements like ``the 
probability distribution of a system at equilibrium is $\exp (-\beta H)$''  were not so interesting.  
The Hamiltonian of $N$ particles could be written as 
\be H=\sum_{i}{p^{2}_{i}\over 2m} +V(x)\ ,
\ee
but the form of $V(x)$ was completely unknown (e.g. in its first two papers Einstein 1902 supposes 
that $V(x)$ is Coulomb like, but he was not able to find a good agreement with the experiments).  
The real interesting problem was the computation of the distribution of the momenta at equilibrium 
and of the related transport coefficients.  Planck made the first successful attempt to go beyond 
kinetic gas theory and to apply statistical mechanics to a completely different systems.

\section{Planck's approach to radiation}

Planck started his carrier as physicist with thermodynamics, a subject that fascinated him. He was
particularly interested to the second law. For 
example, while 
Clausius  usually wrote the second law as:
\be \oint {\delta Q \over T}\le 0\ .
\ee
Planck preferred to use the following mathematical equivalent formulation (which was not unknown to 
Clausius):
\be
S(final)\ge S(initial)\ .
\ee
In this way there was a symmetry among the two laws: the first implies the conservation of energy, 
the second the increase of entropy.

At the beginning Planck had difficult relations with Boltzmann.  He considered Boltzmann's theory 
more or less a useless effort, with not enough results.  When his young assistant Zermelo attacked 
Boltzmann \cite{ZERMELO}, claiming that the Poincar\'e recurrence theorem implies that one cannot 
obtain irreversible effects using a reversible Hamiltonian, he agreed with Zermelo from the 
technical point of view: he seemed not to like that his beloved second law was not exact, but only a 
probabilistic statement.  Contrary to Zermelo, he believed that the second law could be proved by a 
microscopic theory where continuous degrees of freedom were present (later on, he took the point of 
view of Boltzmann on the use of probabilistic methods).  According to Planck, the interaction of 
matter with the radiation is the origine of irreversibility: he started to study the black body 
radiation in order to construct a mechanical proof of the second law.

Why radiation should be the origine of irreversibility?  We know now that the equation of motion of 
matter and radiations are obviously invariant under time reversal, so radiation cannot decide the 
arrow of time.  Which was Planck's approach?  He started \cite{PRIS} by deriving the equation of 
motion of a radiating dipole $f$:
 \be
  {d^{2}f \over dt^{2}}+ \alpha {df \over dt} +\omega^{2}f=E(t)\ ,
 \ee
where the dumping term (proportional to $\alpha$) is due to the emission of radiation and the 
forcing term ($E(t)$) is proportional to the external electric field.  Under this form the equations 
are no more invariant under time reversal, however, in order to derive the previous equation, it is 
necessary to assume the electromagnetic equivalent of Boltzmann molecular chaos (e.g. a random 
distribution of the phases of the different modes in a cavity).  His equation for the dumped 
radiating dipole was in reality derived with the same logical steps of Boltzmann, and when Planck 
realized this, he was able to understand much better Boltzmann's point of view, which he 
publicly joined.

\section{Planck's derivation of his formula for the black body radiation}
 
The main argument of Planck on black body radiation was based on the interaction of the radiation 
field with matter. In his approach,  the simplified model for matter was a set of resonators of 
frequency $\nu$. Resonators are essentially radiation dumped harmonic oscillators (radiating 
dipoles), with the simplification that they can absorb and emit only radiation with the frequency $\nu$.

He firstly studied a resonator with frequency $\nu$.
Using the equations of motion for the dumped harmonic oscillator, he obtained \cite{PWIEN} that
\be
E_{Radiation}(\nu,T)={8 \pi \nu^{2} \over c^{3}} U_{Matter}(\nu,T) \ ,
\ee
where $U_{Matter}(\nu,T)$ is the energy of a resonator of frequency $\nu$ at temperature $T$.  In 
the same framework he had already derived an electromagnetic equivalent of Boltzmann's $H$-theorem.

For Planck the energy $U(\nu,T)$ is fixed by the knowledge of the entropy $S(\nu,U)$, which, 
contrary to any logical argument, is not the usual entropy of the harmonic oscillator.  In other 
words Planck believed that it was consistent to treat the resonators as harmonic oscillators as far 
as the interaction with the radiation was concerned and to suppose that the form of the entropy 
$S(\nu,U)$ was determined by the unknown internal structure of the resonators.  The complete 
ignorance of the structure of matter lead Planck to think that $S(\nu,U)$ could be any function, 
provided that it is in agreement with Wien's displacement law.  Indeed Wien's displacement law is 
satisfied only if
\be
S(\nu,U)=s\left({U \over  \nu }\right) \ .
\ee

Planck believed that $S(\nu,U)$ should be a simple expression. In his first attempt to explain the 
black body  radiation \cite{PWIEN} he obtained the Wien distribution \cite{WDIST}:
\be E(\nu) \propto \nu^{3} \exp(-a \nu/T) \ ,
\ee
assuming that the entropy was given by the expression
\be
S(\nu,U)={-U \over a \nu }\ln{U \over b\nu} \ ,\ee
that seemed particular simple to Planck (he liked the presence of a logarithm).

It may be interesting to notice that the previous equation implies that
\be
{\partial ^{2}S \over \partial  U^{2}} \propto  U^{-1} \label{W} \ ,
\ee
while for a classical harmonic oscillator we have
\be
{\partial ^{2}S \over \partial  U^{2}} \propto  U^{-2} \label{C}\ .
\ee

In the early afternoon of the Sunday October $7^{th}$ of 1900 he learned from Rubens that the data 
for the black body radiation in the far infrared were proportional to the temperature, as follows 
from the classical expression eq.  \ref{C}.  He tried to interpolate among the previous formulae: 
the first was {\sl experimentally} correct at small $U$ and the second was correct at large $U$.  
The simplest interpolation among the two formulae was
\be
{\partial ^{2}S \over \partial  U^{2}} ={-\alpha(\nu) \over U (U+\beta(\nu))} \ .
\ee
Starting from it, he derived in late afternoon of the same day   Planck's law for the black body 
radiation and its expression was sent to Rubens on a postcard the same evening.
Planck's law was presented to the Academy \cite{PDIS1} two weeks later, nearly simultaneously with 
the presentation of the experimental data of Rubens and Kurlbaum.

In the second paper \cite{PDIS2} he computed the entropy of the resonator using the formula
\be
E_{n}= n h \nu \ , \label{Q}
\ee
where this formula applies only to the resonators ({\sl not} to the 
electromagnetic fields).
However he did not assume that  the energy of the resonators is quantized, i.e. it is given by the 
previous formula.
If one reads carefully the paper, one discovers that he did something very different: he  
defined $n$ as the integer nearest to $E/(h \nu)$:
\be
n \approx {E \over h \nu} \ .
\ee
He assumed also that the entropy of the oscillator can be computed by summing over $n$ instead 
of integrating over the energy. All values of the energies are allowed. The quantization of the 
energy enters only in the computation of the entropy \footnote{The formula
$ S= k \ln W $ is written on Boltzmann's grave but it 
was not written by Boltzmann. 
Planck introduced $k$ in the in its first paper on the black body radiation. He was the first to 
write later $S= k \ln W +const $. At the beginning $k$, not $h$, was called Planck constant.}.

\section{The development of statistical mechanics}
 
Nearly  simultaneously to Planck paper, Gibbs and Einstein \cite{GIBBS, ESTAT}, put forward the modern 
statistical mechanics: canonical ensembles, microcanonical ensembles, thermal fluctuations and all 
that. In this way physicists had the theoretical tools needed to apply statistical mechanics to 
many different problems.
 
One of the main consequences of Planck work was the possibility of computing the specific heat of 
solids and of diatomic gases at low temperature.  The first step was done by Einstein in 1908 and it 
was followed by a large number of papers.  The subject was very important and it was a crucial 
ingredient for computing the absolute value of the entropy, which had important practical relevance 
in the study of chemical equilibrium \cite{NERST}.  One could say that the construction of models in 
statistical mechanics started in order to reach a better understanding of the specific heat of 
solids.  In this way statistical mechanics was applied to many different cases and the original 
kinetic gas theory becomes only a restricted field.

However, these new developments could happen only after a clear understanding of Planck's original 
proofs.  It was crucial to establish which was the correct way to apply statistical mechanics to 
black body radiation and to other quantum systems.  There was a strong debate and eventually it was 
clear that  Planck's proofs were inconsistent.  The conclusions was that according to classical 
statistical mechanics the properties of the resonators are irrelevant: the equipartition law should 
be valid for the electromagnetic field, which is a nearly perfect harmonic oscillator, 
irrespectively of any contrived form of the entropy of the resonator.  The entropy of the resonator 
could be extremely complex due to non linear effects, without affecting in any way the black-body 
radiation \cite{ERE}.

It was  also realized that, at thermal equilibrium  in classical mechanics,  
the Rayleigh-Einstein-Jean law \cite{REL,ERJEJ,JEANS} must be valid:
\be
E_{Radiation}(\nu)={8 \pi \nu^{2} \over c^{3}} kT \ ,
\ee 
This law coincide with  Planck's law,
\be
E_{Radiation}(\nu)={8 \pi \nu^{2} \over c^{3}} {h \nu \over \exp (h \nu/kT) -1}\ ,
\ee
only in the region of small frequencies.

It is evident that the Rayleigh-Einstein-Jean law is not experimentally correct, because it corresponds to an 
infinite electromagnetic energy \footnote{The fact  that classical mechanics leads 
to the so called ultraviolet catastrofe was discovered only a few years after Planck's paper and not 
viceversa, as suggested by the  usual folklore.}.
The solution to this paradox was not evident. Rayleigh, Jeans, Einstein and Planck had very different 
opinions on the direction which should be taken.
\subsection{Rayleigh's solution}

The point of view of Rayleigh \cite{JEANS} was that the equipartition law (and consequently 
statistical mechanics) is not correct in general.  It is valid for the low frequency modes, but not 
at high frequencies.  One could paradoxically ascribe this opinion to Boltzmann: he was the first to 
notice that the oscillatory degrees of freedom, that can be observed in the absorption and emission 
spectrum, do not contribute to the specific heat of a solid or of a gas.  Equipartition gave always 
the wrong result at high frequency.  A wrong hypothesis should be present in its derivation: where 
else the error could lurk?

\subsection{Jean's solution} Roughly speaking, Jeans  believed that at thermal equilibrium 
the Rayleigh-Einstein-Jean law should be valid, bur the black body radiation could take an extremely long 
time to reach equilibrium \cite{JEANS}.  In real experiments the radiation should have not yet reached the 
equilibrium. The Planck distribution was only an approximate law, where $h$ could be a time dependent 
decreasing function.

However, if black body radiation is not at equilibrium, the thermodynamical derivation of the Wien 
law (eq.  \ref{WIEN}) was no more valid. People were reluctant to abandon one of the best verified 
law on black body radiation and Jean's proposal was not very successful.

\subsection{Einstein's solution} 
Einstein's solution was the photon, i.e. he supposed that the energy of the electromagnetic field is 
quantized.  In a beautiful paper in 1909 \cite{EPHOT} he observed that if the Wien law were correct, 
the entropy of the electromagnetic field would be the same as that of $N$ independent pointlike quanta of 
energy given by eq. \ref{Q}.  Einstein used this argument for jumping to the correct conclusion that the 
energy of the electromagnetic field is quantized and each of the $N$ quanta (the photons) may be 
absorbed independently from the other (a position he had already taken in the explanation of the 
photoelectric effect in 1905).

As far I understand, Einstein did not realize that in the regime where the Wien law is correct, the average 
value of $N$ is very small and it can be computed assuming that $N$, may takes the values 0 or 1.  
Indeed,  Planck's law does not correspond to a gas of classical non interacting photons.  We know now 
that only in the deep quantum regime the correct expression for the quantum photon gas coincides with 
that of a classical photon gas: the deep quantum regime for the wave is the classical regime for 
particles.

The first modern proof of  Planck law's, where one assumes that the energy of the 
electromagnetic field is quantized (i.e. satisfies eq.  \ref{Q}), appeared in 1910 \cite{DEB}. It is 
interesting to note that many physicists think that this  is the original proof presented by 
Planck.

Einstein's proposal, i.e. the quantization of the electromagnetic field, was treated as heretical: 
in 1911 Planck wrote a letter where he said  he supported Einstein candidature for a chair in 
spite of the fact that he has done the mistake of proposing the photon.  The general opinion was 
that the resonator could behave in a strange way: its equation of motion were not known, however the 
electromagnetic field was described by the Maxwell equations and there was no place for a quantized 
electromagnetic field. The Maxwell equations were in very good agreement with 
experiments and there was no reason to assume that they were not correct. In other words, many 
physicists believed for a long time that quantization was the non-linear effect of a classical dynamics 
and that there was no place for it, if the equations of motions were linear.

\subsection{Planck's solution}
Planck was rejecting the quantization of  electromagnetic energy. It was also rejecting the 
possibility that the resonator could absorb the energy from the electromagnetic field in a discontinuous 
way. He was assuming that the microscopic system were evolving in a continuous way. He could accept 
the quantization in the emission of radiation, as a very fast process in which the radiation goes 
from the resonator to the electromagnetic field; the  process should be so fast that it could be considered 
discontinuous. He could not accept the quantization in the absorption of radiation, because, as an effect 
of  local energy conservation, the absorption  of  a very energetic quantum should take a 
very large amount of time \footnote{The time should exponentially diverge when the energy increases 
because the amount of available energy in the high frequency modes is exponentially small.} .

He put forward the following rather strange theory: the resonators adsorb energy in a continuous way 
from the electromagnetic field and they change their energy in a continuous way.  The resonator may 
emit a radiation equal to $E_{n}\equiv n h \nu$ only when its energy is exactly equal to $E_{n}$ and 
in this way it makes a discontinuous transition to the state of zero energy.

This proposal has nothing to do with  reality, but it contains many new ideas which had a deep 
influence on later developments:
\begin{itemize}
    \item The different orbits ($H=E_{n}$) in the two dimensional space divide the  phase space in 
    regions of area $h$. We could call each region a quantum state. In this way, each  quantum state 
    occupies an area in phase space equal to $h$:  in this respect $h$ appears as  a quantum of 
    action. The similarities at technical level with the later work of Bohr and Sommerfeld are striking.
    \item The average energy of a quantum state is 
    \be
    {E_{n}+E_{n+1}\over 2}=(n+\frac12) h \nu \ .
    \ee 
    The minimal energy of a resonator is $\frac12 h \nu $.  The zero point energy enters in physics 
    for the first time \footnote{It is amusing to note that the proposal of a zero point energy 
    soon became  popular: it was  applied it to the specific heat of diatomic gas, showing that 
    the experimental data presented a strong support for a zero point energy in the rotational 
    spectrum \cite{ZERO} (nowadays we know that zero point energy in the rotational spectrum is 
    not present: in the $s$-wave the rotational energy is zero!)}.  
    \item The emission of the 
    radiation (when the oscillator crosses the level $E_{n}$) is a probabilistic process, which 
    happens with a given probability.  This was the first case where a transition 
    probability between quantum states was introduced .  
    Planck's intuition that the emission of radiation should 
    not be a deterministic process is remarkable.
\end{itemize}

\section{Quantum Statistical Mechanics}

Of course, the final solution of the previous dilemma was quantum statistical mechanics: Einstein was 
right and the energies of both matter and the electromagnetic fields are quantized (in a first 
approximation according to the Bohr-Sommerfeld quantization formula).  This solution took long time 
to be completely accepted.  Still in 1924 Bohr did not believed in the reality of the photon.  
According to him, the electrons were jumping in discontinuous way, but they were emitting and 
adsorbing energy in continuous way; this was possible due to the fact that energy and momentum were 
not conserved in individual quantum events \cite{BOHR}.  The experimental observation of the Comtpon 
effect, where it was possible to check the conservation of energy and momentum in a given 
individual process, was the definitive and needed proof of the quantization of the energy of the 
electromagnetic field and it convinced nearly everybody.

However, quantum statistical mechanics for many particles was not easy to establish. The Planck 
formula did not correspond to a gas of classical photons. 

The way to attack this problem was found by Bose in the year 1924 \cite{BOSE}.  According to classical 
mechanics, if we have $N$ particles which can stay in three different states and we denote with 
$k_{i}$ the number of particles in the $i^{th}$state ($k_{1}+k_{2}+k_{3}=N$), the probability has a 
combinatorial prefactor which is given by:
\be
P(k_{1},k_{2},k_{3})={
N! \over k_{1}!k_{2}!k_{3}!}\ , \label{CLASSICA}
\ee
i.e. by the number of combinations in which the $N$ particles may be divided in three sets. This 
formula is sometimes called the Boltzmann counting, because it was used by Boltzmann.

According to quantum mechanics, if the particles are indistinguishable and they are Bosons, the wave 
function is symmetric and there is only one quantum state for each set of occupation numbers.  
Therefore in this case the correct formula is simply given by:
\be
P(k_{1},k_{2},k_{3})=1 \ . \label{CORRECT}
\ee

The way in which this formula was found is remarkable.  Bose started by introducing the quantity 
$n_{k}$ that are equal to the number of states with $k$ particles (i.e. $\sum_{i} 
\delta_{k,k_{i}}=n_{k}$)\footnote{For example in the case where we have 4 states, 8 particles and 
$k_{1}=3,\ k_{2}=2,\ k_{3}=2,\ k_{4}=1$, we have $n_{3}=n_{1}=1,\ n_{2}=2,\ 
n_{0}=0$.  One obviously has that $\sum_{k} k \ n_{k}= N$.}.  The main step by Bose was to suppose 
that
\be
P(n_{0}, n_{1}, n_{2}) \propto {N! \over n_{0}! n_{1}! n_{2}! \ldots} \ . \label{MAGICA}
\ee

The amazing point is that Bose did not give any explanation for his formula (which has no classical 
justification) and he declared that it is evident!  According to Pais \cite{KP} it is reasonable to suppose 
that Bose has written eq.  \ref{MAGICA} in analogy with the classical Boltzmann counting (eq.  
\ref{CLASSICA}), without realizing that the $n$'s and the $k$'s are very different objects.  In other 
words, his formula was based on a misunderstanding and his derivation was not correct. On the other 
hand, it was impossible to derive the Bose formula (eq.  \ref{MAGICA}) from what was known at that time.

Only two years later, Fermi extended the argument to the case of Fermions, and Dirac presented a 
microscopical derivation of both statistics from the assumptions of symmetry (or antisymmetry) of 
the wave functions. He was also the first to point out that the mysterious Bose's formula (eq.  
\ref{MAGICA}) could have been written in a simple way as eq.  \ref{CORRECT}.

\section{Were Rayleigh and Jeans correct?}

Statistical mechanics developed  at a steady pace for the rest of the century. 

One of the main 
problems, the existence of phase transition in the canonical ensemble was a puzzle for a long time. 
In a finite volume everything (e.g. the energy) is an analytic function of the temperature and 
a  phase transition cannot exist.

The modern solution (i.e. a phase transition is present only in the infinite volume limit) appeared quite 
slowly: at the van der Waals conference (1937) there was a public discussion and 50\% of the 
physicists expressed the opinion that a phase transition cannot be present in a formalism based on 
the partition function \cite{WDW}.  Fortunately the situation changed very fast; in 1943 Onsager 
obtained its solution of the two dimensional Ising model, where a phase transition is present, and 
in 1951 Lee and Yang presented a clear analysis of the mathematical mechanism at the basis of the 
phase transitions.

In the later years we had good and bad news on the foundation of statistical mechanics.

The good new was that
Sinai proved that for a small number of hard spheres the microcanonical distribution is correct 
(actually  he proved much more).

The bad new was that Fermi, Pasta and Ulam took a chain of anharmonic oscillators, studied their 
evolution on a computer and found that the microcanonical distribution is not reached. The 
Hamiltonian they were considering was
\be
H=\sum_{i=1,N}\left({p_{i}^{2}\over 2}+  {(\phi_{i}-\phi_{i+1})^{2}\over 2 }
+g {(\phi_{i}-\phi_{i+1})^{4}\over 4}\right),
\ee
 The corresponding equation of motions in the continuum limit are 
 \be
 {\partial ^{2} \phi \over \partial t^{2}}={\partial ^{2} \phi \over \partial x^{2}}+g
 {\partial \over \partial x}\left({\partial  \phi \over \partial x}\right)^{3}
 \ee

At the same time KAM theorem was proved. Very roughly speaking, KAM theorem states  that in a 
systems of $N$ weakly coupled   oscillators, when the anharmonicity $g$ is small, there are $N$ constant 
of motions, in the same way as in the uncoupled case at $g=0$. Only for large values of $g$ the 
energy (and the momentum) is the unique conserved quantity.

Both papers show that  for finite $N$ and not too large anharmonicity Rayleigh was correct and 
equipartition law is not satisfied.
Many numerical simulations  \cite{GALGANI} imply that
 for finite $N$ there is an ergodicity threshold in $g$.
 
A relevant question is to establish what happens for $N$ going to infinity. One can present 
hand waving arguments,  suggesting that in the limit $N \to \infty$ at non-zero $g$ the system will 
reach thermal equilibrium and an upper limit on the equilibration time ($\tau(g)$) is proportional to 
$\exp(A/g)$; for times much larger that $\tau(g)$ an infinite system is at thermal equilibrium 
\cite{PARISI1}.
 
The conclusion is that in the case of a generic weakly anharmonic system Jeans was correct: 
the equipartition law is correct, but the time to reach equilibrium may be very large.  This effect, 
 due to the difficulty in transferring energy from the low frequency modes to the high 
frequency modes, could be observed experimentally (not only in numerical simulations) by heating a 
crystal by pumping the acoustic phonons and looking for the energy transfer to optical phonons.

An interesting and open question  is the precise behaviour of $\tau(g)$ for small $g$. There are no 
doubts that $\tau(g)$ is strongly divergent for small $g$, however its precise form is not evident.
A careful numerical experiment has been done for the slightly different Hamiltonian \cite{PARISI2}:

\be
H=\sum_{i=1,N}\left({p_{i}^{2}\over 2}+{\phi_{i}^{2}\over 2} +{(\phi_{i}-\phi_{i+1})^{2}\over 2 }
+g {\phi_{i}^{4}\over 4}\right)\ .
\ee
The corresponding equation of motion in the continuum limit are
\be
{\partial ^{2} \phi\over \partial t^{2}}={\partial ^{2} \phi\over \partial x^{2}}+ m^{2}  \phi +g 
\phi(x)^{3}\ .
\ee

In this case  the numerical data suggest a behaviour of the form
\be \tau (g)\propto g^{-4}\ee \ .
In 
the original Fermi Pasta Ulam model other numerical simulations suggest \cite{FPU2}  the 
behaviour $\tau (g)\propto g^{-3}$.

Which is the correct behaviour? Numerical simulations may hint in some directions but they cannot be 
conclusive if a theoretical understanding is lacking.
If the behaviour of $\tau(g)$ is a power law (as it looks like) it should be possible to control it in 
perturbation theory. Unfortunately such an analysis has not be done and 
the answer is not known.

\section{Conclusions}

In my opinion one of the most impressive fact is the large number of wrong statements, 
wrong beliefs and incomplete proofs that can be found in these historical papers. Let me summarize 
some of the 
most impressive:
\begin{itemize}
    \item Planck believed that the interaction of matter with the radiation was the key to explain 
    irreversibility without using molecular chaos.
    \item
    Planck did not realize that at equilibrium in classical mechanics the properties of the resonators 
    have no influence on the black body radiation.
    \item Einstein deduced the existence of the photon from the properties of the black body 
    radiation showing that Wien's law implies that the radiation behaves as $N$ independent {\sl photons}
    without noticing that Planck's law implies that as soon $N>1$ (i.e. in the whole non trivial 
    region) the photons are not independent.
    \item
    Bose wrote his main formula stating that it was evident and it did not realize that there were no 
    reasons for its validity.
\end{itemize}
It is quite likely that quantum mechanics could be discovered only in this illogical way.  The gap 
among classical and quantum mechanics is too large to be crossed in one step only.  If people were 
consistent and correct at all time, they would never move and they would make no progress, because 
their aim, to find a classical explanation of quantum phenomena, was an impossible task.  For 
example, it is not surprising that Planck presented a wrong classical proof of the black body 
radiation formula, because a correct proof of this statement does not exist.  The influence of 
classical mechanics was so strong that at the beginning people always tried to wrongly derive 
quantum formulae in a classical contest.

These great men had the ability (conscious or unconscious, who knows?)  to isolate the physical 
relevant points, where progress could be done, not paying attention to the possible contradictions, 
not exploring in their first papers the dangerous points where inconsistencies may arise.  Only 
after that the experimental correctness of the theory was verified, these points could be discussed 
without destroying the theory.

It is difficult to predict what could have been the history of physics if Planck had make the 
correct classical equilibrium statement, i.e. the Raleigh-Einstein-Jeans law.  As stated by 
Einstein, it is quite likely that Planck would not have discovered Planck's law.  It is possible 
that the efforts would have concentrated on one (or both) of the directions suggested by Raleigh and 
Jeans.  In this case, physicists would have tried to solve a problem, which a century later we have 
not solved, moreover without having any ideas of the forces involved.  In spite of the difficulties 
this program could have been partially successfully and when Planck's law would be later discovered 
it could be considered a nice fit, to be explained in a non-quantum context.  It is likely that 
quantum mechanics would eventually be discovered, but it would have been much more difficult.

It seems to me that science advances by small steps and therefore, if we have to construct a 
completely new theory, at some stage we must go through inconsistencies.  Only great men, that have 
a strong physical intuition for what should be right and what should be wrong, could successfully 
navigate in these dangerous waters.

\end{document}